\documentclass[authoryear, 5p]{elsarticle}
\usepackage[colorlinks]{hyperref}
\usepackage[latin1]{inputenc}       
\usepackage[T1]{fontenc}            
\usepackage{graphicx}        
\usepackage{multicol}               
\usepackage{multirow}
\usepackage{booktabs}               
\begin{document}
\begin{frontmatter}
\title{\small This paper has been published in Environmental Modelling \& Software, 2010, 25, 1479-1480, \href{http://dx.doi.org/10.1016/j.envsoft.2009.01.003}{doi:10.1016/j.envsoft.2009.01.003}\\[0.5cm] \LARGE EcoTRADE - a multi player network game of a tradable permit market for biodiversity credits \\}
\journal{Environmental Modelling and Software}
\author[UFZ]{Florian Hartig\corref{cor}}
\cortext[cor]{Corresponding author, Tel: +49-341-235-1716, Fax: +49-341-235-1473}
\ead{florian.hartig@ufz.de}
\author[U-Leipzig]{Martin Horn}
\ead{martin\_horn@gmx.de}
\author[UFZ]{Martin Drechsler}
\ead{martin.drechsler@ufz.de}

\address[UFZ]{UFZ - Helmholtz Centre for Environmental Research, Department of Ecological Modelling, Permoserstr. 15, 04318 Leipzig, Germany}
\address[U-Leipzig]{University of Leipzig, Department of Computer Science, Johannisgasse 26, 04103 Leipzig, Germany}
\begin{abstract}
EcoTRADE is a multi player network game of a virtual biodiversity credit market. Each player controls the land use of a certain amount of parcels on a virtual landscape. The biodiversity credits of a particular parcel depend on neighboring parcels, which may be owned by other players. The game can be used to study the strategies of players in experiments or classroom games and also as a communication tool for stakeholders participating in credit markets that include spatially interdependent credits.
\end{abstract}

\end{frontmatter}

\section{Software availability} Name of the software: EcoTRADE \\
Availability: Software, documentation and an online applet are available at \\ \href{http://www.ecotrade.ufz.de/ecotradegame.html}{www.ecotrade.ufz.de/ecotradegame.html}\\
Developers: Martin Horn, Martin Drechsler, Florian Hartig \\
Year first available: 2008 \\
Software required: Java (JRE version 1.5 and higher). Additionally, the web browser must allow the execution of java applets for an optional online view. \\
Programming language: Java \\

\section{Introduction}
Tradable permits are an economic instrument for controlling the use of environmental resources. Examples of tradable permits include the carbon emissions trading schemes settled under the Kyoto Protocol or the tradable permit system for restricting emissions of ozone-depleting chemicals in the US after 1988 \citep{Tietenberg-EmissionsTradingPrinciples-2006}. In recent years, tradable permit schemes with names such as biobanking or biodiversity credit trading have also been applied to restrict land use and ensure the maintenance of natural habitats and biodiversity \citep{Wissel-ConceptualAnalysisof-2010}. Yet, unlike in the case of carbon emissions, biodiversity credits can not be issued independently of the spatial location. Species depend on the connectedness of their habitat. Therefore, the ecological benefit of conserving a site is higher in the vicinity of other conserved sites and credits should be issued accordingly \citep{Drechsler-Applyingtradablepermits-2009, Hartig-Smartspatialincentives-2009}.

\begin{figure*}[htb]
\centering
\includegraphics [width=14cm]{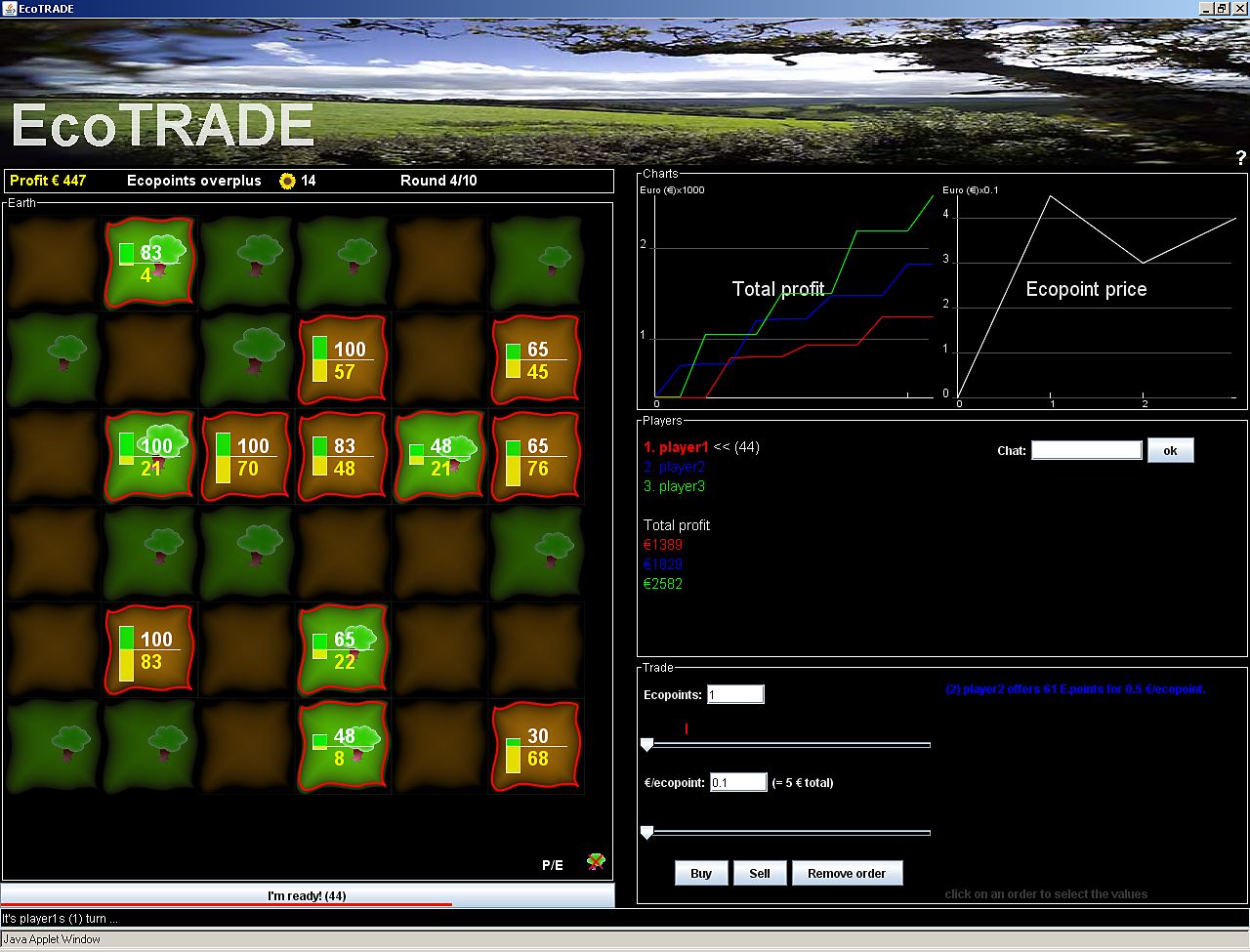} 
\caption{A three-player game, showing the client window of player~1. On the left hand side we see the game landscape. Properties belonging to player~1 are highlighted. Each of these properties is labelled with two numbers; the upper one denotes the conservation credits earned from the property if used for conservation, the lower one marks the economic profit earned from the property if used for agriculture. A tree on a property indicates that the site is used for conservation. On the right hand side (from top to bottom), we see the current profits of all players and the time series of prices of ecopoint transactions (top); status information and a chat window (middle); and a window for credit trading (bottom).}
\label{figure: ecotrade client}
\end{figure*} 

The multiplayer game EcoTRADE illustrates such a situation: In a virtual biodiversity credit market, players may use their land for conservation or agricultural purposes. The amount of biodiversity credits they receive depends on the land use in their neighborhood. Players interact through these neighborhood-dependent benefits and by trading credits on a virtual market. The software may be used for educational purposes, such as classroom experiments, as well as for communication with stakeholders or for experimental economics.  
  
\section{Game description and software features}
At the beginning of the game, each player receives a certain land entitlement and a certain conservation obligation. To produce more or less biodiversity credits, players can change the land use type of a parcel through mouse clicks. If players want to undersupply their obligation, they first need to buy credits from other players. An overplus of credits can be sold to other players. The aim of the game is to maximize economic profits from the land, which can be done by collecting revenue from agriculturally used fields or by selling biodiversity credits to other players (see Fig.~\ref{figure: ecotrade client}).

The software consists of a server and a client. The clients, controlled by the players, may connect to the server through any network connection, i.e. via a local network or the internet. Players may initialize new games on the server, modify the rules of the new game, and invite other players to join in. All parts of the software are written in Java and can be run on any system which provides a java virtual machine. User manuals in English and German can be found on the game website, from where it is also possible to play the game directly in the web browser. The software also contains an experimental one-player version, where the second player is controlled by the computer.

\section{Practical experience and concluding remarks}
Credit markets with spatially dependent credits give rise to a number of questions. Landowners may influence each other's payoffs \citep[see][]{Hartig-Staybythy-2010}. Will they coordinate or cooperate to optimize their profits? Will they be able to find the land configurations that optimize their payoffs? Will players find it unfair that other players may influence their payoffs? 

The EcoTRADE software has been used at several workshops with participants ranging from children between the ages of 10 and 15 years to ecology and economics students and scientists working in the field of conservation and resource economics. We found that the software is useful for observing the typical behavior and strategies of players, but also for communicating problems arising from spatially interdependent credits. In conclusion, we hope that users will find the EcoTRADE game useful as a communication tool for classroom games and stakeholder workshops, but also for experimental studies in economics.


\begin{thebibliography}{5}
\expandafter\ifx\csname natexlab\endcsname\relax\def\natexlab#1{#1}\fi

\bibitem[{Drechsler and Wätzold(2009)}]{Drechsler-Applyingtradablepermits-2009}
Drechsler, M., Wätzold, F., 2009. Applying tradable permits to biodiversity
  conservation: Effects of space-dependent conservation benefits and cost
  heterogeneity on habitat allocation.
  \href{http://dx.doi.org/doi:10.1016/j.ecolecon.2008.07.019}{Ecological
  Economics} 68~(4), 1083--1092.
\newline doi: doi:10.1016/j.ecolecon.2008.07.019

\bibitem[{Hartig and Drechsler(2009)}]{Hartig-Smartspatialincentives-2009}
Hartig, F., Drechsler, M., 2009. Smart spatial incentives for market-based
  conservation.
  \href{http://dx.doi.org/10.1016/j.biocon.2008.12.014}{Biological
  Conservation} 142~(4), 779--788,
  \href{http://arxiv.org/abs/0809.0228}{http://arxiv.org/abs/0809.0228}.
\newline doi: 10.1016/j.biocon.2008.12.014

\bibitem[{Hartig and Drechsler(2010)}]{Hartig-Staybythy-2010}
Hartig, F., Drechsler, M., 2010. Stay by thy neighbor? Social organization
  determines the efficiency of biodiversity markets with spatial incentives.
  \href{http://dx.doi.org/10.1016/j.ecocom.2009.07.001}{Ecological Complexity}
  7~(1), 91--99.
\newline doi: 10.1016/j.ecocom.2009.07.001

\bibitem[{Tietenberg(2006)}]{Tietenberg-EmissionsTradingPrinciples-2006}
Tietenberg, T., 2006. Emissions Trading Principles and Practice, 2nd Edition.
  RFF Press.

\bibitem[{Wissel and Wätzold(2010)}]{Wissel-ConceptualAnalysisof-2010}
Wissel, S., Wätzold, F., 2010. A Conceptual Analysis of the Application of
  Tradable Permits to Biodiversity Conservation.
  \href{http://dx.doi.org/10.1111/j.1523-1739.2009.01444.x}{Conservation
  Biology} 24~(2), 404--411.
\newline doi: 10.1111/j.1523-1739.2009.01444.x

\end{thebibliography}
\end{document}